\documentclass[preprint,12pt]{elsarticle}
\usepackage[pdfborder={0 0 0}]{hyperref}
\usepackage{amsthm,amsmath,amsfonts,latexsym,amssymb}
\usepackage{graphics,fancyhdr,graphicx,subfigure}
\usepackage[ruled,linesnumbered]{algorithm2e}
\usepackage[margin=1in]{geometry}
\usepackage{multirow,booktabs}
\usepackage[table]{xcolor}
\usepackage{listings}
\usepackage{tabularx}
\usepackage{placeins}
\usepackage{comment}
\usepackage{lineno}
\usepackage{lscape}
\usepackage{bm}
\usepackage{hyperref}
\usepackage{mathrsfs}
\usepackage{arydshln}
\usepackage{appendix}
\definecolor{custom-blue}{RGB}{3,69,173}
\hypersetup{
    colorlinks = true,
    urlcolor   = blue,
    citecolor  = custom-blue,
}

\biboptions{numbers,sort&compress}
\definecolor{listinggray}{gray}{0.9}
\definecolor{lbcolor}{rgb}{0.9,0.9,0.9}
\definecolor{Darkgreen}{RGB}{0,100,0}
\usepackage{diagbox}
\usepackage{multirow}
\usepackage{rotating}

\newtheorem{theorem}{Theorem}

\makeatletter
\def\ps@pprintTitle{%
 \let\@oddhead\@empty
 \let\@evenhead\@empty
 \def\@oddfoot{}%
 \let\@evenfoot\@oddfoot}
\makeatother

\begin{document}
\abovedisplayskip=6.0pt
\belowdisplayskip=6.0pt
\begin{frontmatter}

\title{Sound propagation in realistic interactive 3D scenes with parameterized sources using deep neural operators}

\author[1]{Nikolas Borrel-Jensen}
\author[3]{Somdatta Goswami}
\author[2]{Allan P. Engsig-Karup}
\author[3,4]{George Em Karniadakis}
\author[1]{Cheol-Ho Jeong \corref{cor1}}

\address[1]{Department of Electrical and Photonics Engineering, Acoustic Technology, Technical University of Denmark, Ørsteds Plads, 2800 Kgs. Lyngby, Denmark}
\address[2]{Department of Applied Mathematics and Computer Science, Technical University of Denmark, Richard Petersens Plads, 2800 Kgs. Lyngby, Denmark}
\address[3]{Division of Applied Mathematics, Brown University, 170 Hope Street, Providence, RI - 02906, U.S.A.}
\address[4]{School of Engineering, Brown University, 170 Hope Street, Providence, RI - 02906, U.S.A.}
\cortext[cor1]{Corresponding author.}

\begin{abstract}
\noindent
We address the challenge of acoustic simulations in $3$D virtual rooms with parametric source positions, which have applications in virtual/augmented reality, game audio, and spatial computing. The wave equation can fully describe wave phenomena such as diffraction and interference. However, conventional numerical discretization methods are computationally expensive when simulating hundreds of source and receiver positions, making simulations with parametric source positions impractical. To overcome this limitation, we propose using deep operator networks to approximate linear wave-equation operators. This enables the rapid prediction of sound propagation in realistic $3$D acoustic scenes with parametric source positions, achieving millisecond-scale computations. By learning a compact surrogate model, we avoid the offline calculation and storage of impulse responses for all relevant source/listener pairs. Our experiments, including various complex scene geometries, show good agreement with reference solutions, with root mean squared errors ranging from $0.02$ Pa to $0.10$ Pa. Notably, our method signifies a paradigm shift as -- to our knowledge -- no prior machine learning approach has achieved precise predictions of complete wave fields within realistic domains.
\end{abstract}

\end{frontmatter}


\section{Introduction} \label{sec:intro}
Wave phenomena are precisely described by solving partial differential equations (PDEs) with their approximate solutions found using numerical methods. Many methods exist, such as finite-difference time-domain methods (FDTD) \cite{Botteldoorena1995}, finite-volume time-domain methods (FVTD) \cite{Bilbao2013}, finite/spectral element methods (SEM) \cite{Pind2019}, discontinuous Galerkin methods (DG-FEM) \cite{Melander2020}, boundary element methods (BEM) \cite{Kirkup2007Bem}, and pseudo-spectral Fourier methods \cite{HornikxPSTD2010}, and are all part of the standard toolbox used to successfully solve a variety of real-world physical problems over the last decades. Determining which method to use depends on the nature and difficulty of the problem, geometric complexity, and trade-offs between accuracy and efficiency. However, all these methods require recalculating solutions for different conditions, including initial and boundary conditions, geometry, and specified source and receiver positions. Obtaining a solution even on a $2$D domain is often computationally expensive; hence, solving parameterized PDEs involving multiple parameters or varying conditions can quickly get intractable.

\begin{figure}[!ht]
    \centering    
    \includegraphics[width=\textwidth]{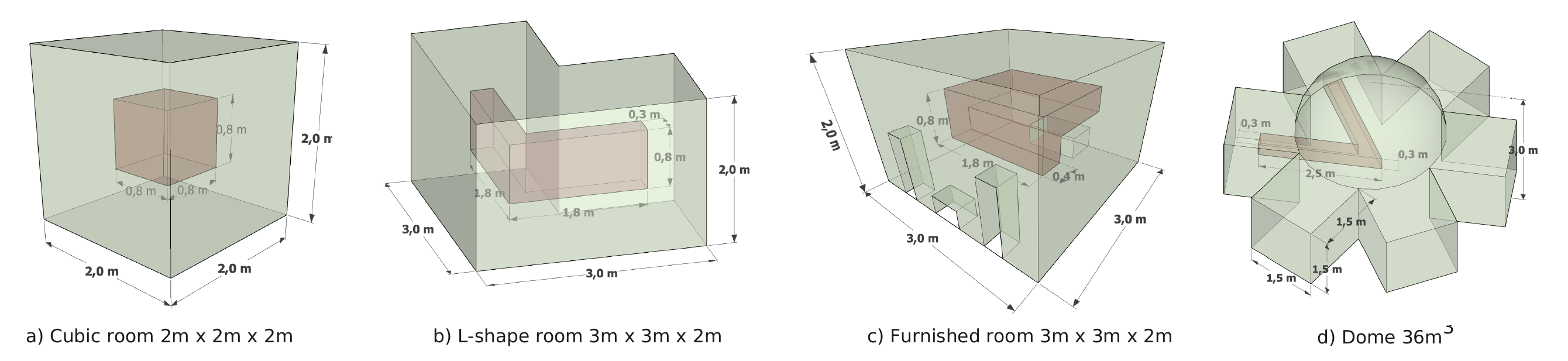}
    \caption{Pictorial representations of the domain geometries adopted in this work to evaluate the predicted $3$D sound fields. All the experiments have parametric source positions allowed to move freely inside a sub-domain of the room shown in shaded red.}\label{fig:geometries}
\end{figure}

In this work, we address the challenges in solving the wave equation for the four geometries depicted in \autoref{fig:geometries} considering its relevance in virtual acoustics, which plays a pivotal role in computer games, mixed reality, and spatial computing \cite{Greenwold2003}. Creating a realistic auditory environment in these applications is crucial for an immersive user experience. The impulse responses (IR) characterizing the room's acoustical properties for a source/receiver pair can be obtained using the numerical methods referenced at the beginning of the section. This is done offline for real-time applications due to the computational requirements, especially when spanning a broad frequency range. However, for dynamic, interactive scenes with numerous parametric source and receiver pairs, the storage requirement for a lookup database becomes intractable (in the gigabytes range). These challenges become even more extensive when covering the audible frequency range up to $20$ kHz. Employing surrogate models to learn the parametrized solutions to the wave equation to obtain a one-shot continuous wave propagation in interactive scenes \cite{nborrel2021} offers an ideal framework to address the prevailing challenges in virtual acoustics applications, effectively surpassing the limitations of traditional numerical methods.

The idea of approximating continuous nonlinear operators for parametrized PDEs from labeled data was first introduced in $1995$ by Chen \& Chen \cite{Chen1995} providing a universal operator approximation theorem for shallow neural networks, guaranteeing small approximation errors (the error between the target operator and the predictions from a class of infinitely wide neural network architectures). Recently in $2019$, Lu et al. \cite{Lu2021} reformulated Chen \& Chen's theorem and generalized the work by proposing the deep operator network architecture `DeepONet,' which exhibits small generalization errors (the ability of a neural network to produce small errors for unseen data). Acknowledging the previous successful application of DeepONet in fracture mechanics \cite{goswami2022physics}, diesel engine \cite{kumar2023real},  microstructure evolution \cite{oommen2022learning}, bubble dynamics \cite{lin2021operator}, bio-mechanics to detect aortic aneurysm \cite{goswami2022neural} and airfoil shape optimization \cite{shukla2023deep}, to name a few, we consider this to be a suitable candidate for our problem.

Despite being a simple non-stiff second-order linear hyperbolic PDE, solving the wave equation is still challenging due to its multi-modal broadband-frequency nature. Therefore, learning a compact and efficient surrogate model to approximate the continuous operators of the wave equation emerges as a valuable solution for addressing a significant real-world challenge, such as virtual acoustics. The resulting DeepONet-based surrogate model should then: $1$) predict the wave field propagation in rooms with parameterized sources and realistic frequency-dependent sources; $2$) produce sufficiently accurate predictions for intended applications; and finally, $3$) infer in real-time ($< 100\text{ ms}$). However, the predictive performance of DeepONet is often restricted by the availability of high-fidelity labeled datasets used for training. Moreover, undertaking isolated learning, which involves training a single predictive model for different yet related single tasks, can be exceedingly expensive. To mitigate this bottleneck, we have introduced a simple transfer-learning framework to transfer knowledge between relevant domains \cite{goswami2022deep}. The transfer learning framework allows the target model to be trained with limited labeled data to approximate solutions on a different but related domain, achieving the same accuracy as the source model, a model trained with a sufficiently labeled dataset on a specific domain. Finally, we push the boundaries of the seminal DeepONet to propose a domain-decomposition framework, which leverages the inherent property of deploying multiple deep neural networks in smaller subdomains, allowing for parallelization. Additionally, it is designed to handle large complex geometries, further expanding the applicability and scalability of the DeepONet method.

\section{Results}\label{sec:results}
Four geometries, in increasing order of complexity, depicted in \autoref{fig:geometries}, have been considered to evaluate the predicted $3$D sound fields in $a)$ a cubic $2\text{ m}\times2\text{ m}\times2\text{ m}$ room with frequency-dependent boundaries, $b)$ an L-shape room with outer dimensions $3\text{ m}\times3\text{ m}\times2\text{ m}$ and frequency-dependent boundaries, $c)$ a furnished room $3\text{ m}\times3\text{ m}\times2\text{ m}$ with frequency-dependent walls, ceiling and floor, and frequency-independent furniture, and $d)$ a dome with a volume of $36\text{ m}^3$ consisting of frequency-independent boundaries. For all the geometries, the models are learned through a final simulation time $T=0.05$ seconds with parametric source positions allowed to move freely inside a sub-domain of the room shown in shaded red. The simulation time was chosen long enough to capture enough information to be meaningful and small enough to make the data generation and training time tractable. The impulse response consists of a direct sound followed by early reflections, which plays a key role in sound perception in rooms up to about 50-100 ms \cite{kuttruff2016room, ISO3382-1} -- slightly above the simulation time in this work. After the sound propagates over time, the response approaches decaying Gaussian noise, referred to as late reverberation. This part is known to be less crucial in sound perception and could be approximated by some statistical method \cite{Raghuvanshi2014}.

Two experiments for the dome are performed; one where the model is trained for receiver positions in the full domain and another where the model is trained for receiver positions in $1/4$ of the domain (denoted `quarter model' in the rest of the manuscript); both cases allow for the source to move freely in the same subdomain. The quarter model applies a domain decomposition approach where separate DeepONets are trained on individual partitions for improved accuracy. 

The training data has been generated using $\texttt{ppw} = 6$ points per wavelength, whereas validation and testing data has been generated using $\texttt{ppw} = 5$ to ensure (mostly) non-overlapping spatial samples to investigate the model's generalization capabilities; \textit{i.e.,} how well the network interpolates at the receiver position, which is crucial for the applications of interest. The input function denoting a Gaussian pulse acting as an initial condition (\autoref{eq:initial_cond}) is sampled at the Nyquist limit, whereas the density of the source positions is sampled at one-fifth of a wavelength for the training data, one full wavelength for the validation data, and five positions for the test data. Details about the data set and DeepONet network setup can be found in Materials and Methods.
The data set sizes are summarized in \autoref{tab:dataset_sizes} ranging from $5.8\text{M} - 21.5\text{M}$ training samples, depending on the complexity of the geometry. The testing data is generated on the same grid as the validation data (different from the training data grid) but only for the five source/receiver pairs. Representative plots of the wave field reference and the corresponding error for the four geometries are presented in Figures \ref{fig:prediction_3D_results_1}-\ref{fig:prediction_3D_results_4}. The plots also present the reference and prediction for the impulse response and the transfer function shown for each source/receiver pair. In \autoref{tab:pred_errors}, the root mean square error (RMSE) for the IR is reported after performing $50-70$k iterations until saturation (\autoref{fig:convergence_plot}).

\subsection{Cubic room}
\begin{figure}[t]
    \centering    
    \includegraphics[width=\textwidth]{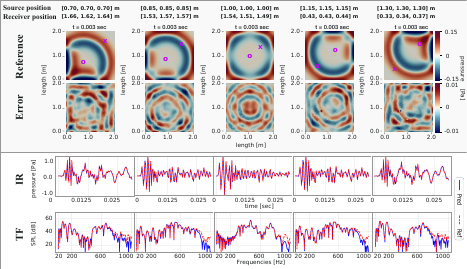}
    \caption{Cubic room $2\times2\times2$ m$^3$. Results show the sound field at $t = 0.003$ s for five parameterized source positions. The wave field error is depicted in the second row, and the IRs and TFs references and predictions are at the two bottom rows. `o'=source position, `x'=receiver position.}\label{fig:prediction_3D_results_1}    
\end{figure}
\autoref{fig:prediction_3D_results_1} shows an almost perfect fit between references and predictions and only minor differences in the upper-frequency range with a mean broadband RMSE of $0.03$ Pa.

\subsection{L-shape room}
\begin{figure}[t]    
    \includegraphics[width=\textwidth]{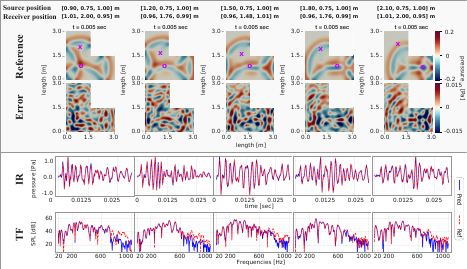} 
    \caption{L-shape room with outer dimension $3\times3\times2$ m$^3$. Results show the sound field at $t = 0.005$ s for five parameterized source positions. The wave field error is depicted in the second row, and the IRs and TFs references and predictions are at the two bottom rows. `o'=source position, `x'=receiver position.}
    \label{fig:prediction_3D_results_2}
\end{figure}
Similar to the previous example, in \autoref{fig:prediction_3D_results_2}, we see a good match between reference and prediction but with bigger deviations in the upper-frequency range above the $700$--$800$ Hz limit. This deviation is also reflected in the mean RMSE of $0.05$ Pa.

\subsection{Furnished room}
\begin{figure}[t]    
    \includegraphics[width=\textwidth]{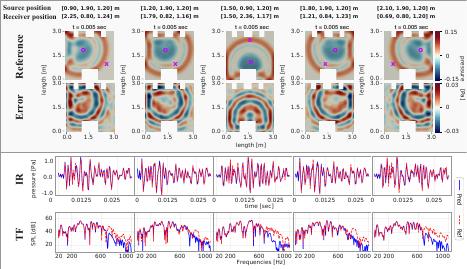}
    \caption{Furnished room $3\times3\times2$ m$^3$. Results show the sound field at $t = 0.005$ s for five parameterized source positions. The wave field error is depicted in the second row, and the IRs and TFs references and predictions are at the two bottom rows. `o'=source position, `x'=receiver position.}  
    \label{fig:prediction_3D_results_3}
\end{figure}
As shown in \autoref{fig:prediction_3D_results_3}, the wave propagation is well captured with quite good agreement between reference and prediction. Still, some inaccuracies are lacking for the sharp peaks, which can also be seen in the upper-frequency range above $600$--$700$ Hz. The mean RMSE is $0.09$ Pa, almost twice the error compared to the L-shape room and three times the error compared to the Cubic room.

\subsection{Dome}
\begin{figure}[t]
    \includegraphics[width=\textwidth]{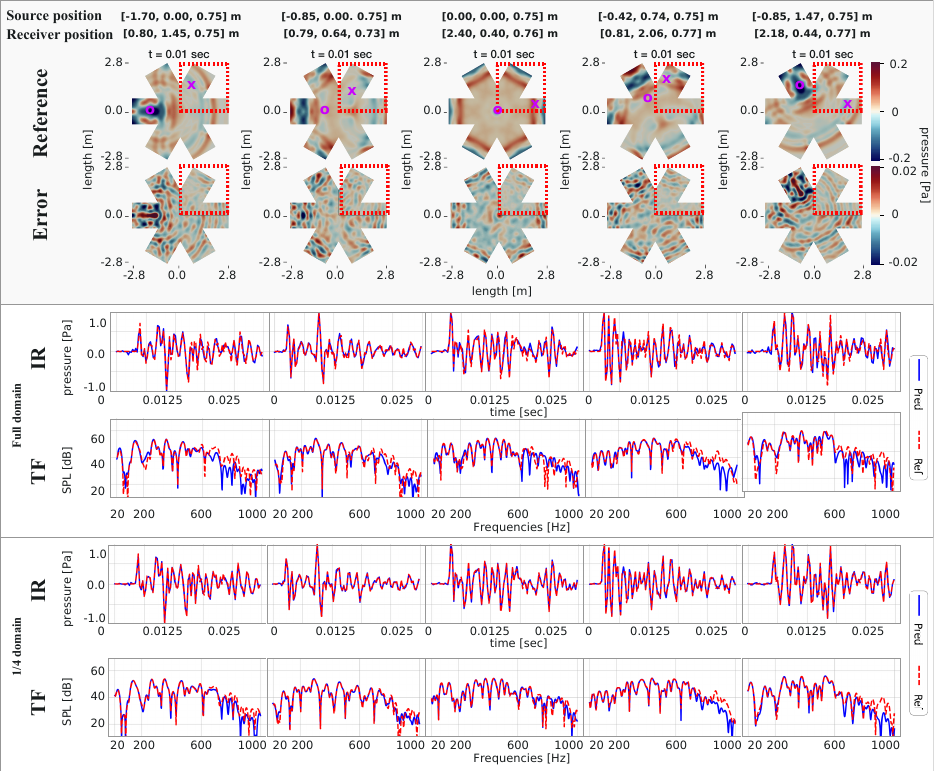}
    \caption{Dome $36\text{ m}^3$. Results show the sound field at $t = 0.01$ s for five parameterized source positions. The IRs and TFs references and predictions are at the two bottom rows for the full and quarter partition. The red square denotes the receiver positions where the quarter model was trained.} 
    \label{fig:prediction_3D_results_4}
\end{figure}
The results for both the full and quarter models are evaluated at the same source and receiver positions for comparison shown in \autoref{fig:prediction_3D_results_4}. The receiver positions are restricted to the $1^{\text{st}}$ quadrant where the quarter model was trained (denoted by the red square) and evaluated at five source/receiver pairs. The wave propagation is well captured for both the full and quarter models, with good agreement between the reference and the prediction. However, not all sharp peaks are well captured for the full model. The fit in the frequency domain is better than the furnished room but not quite as good as for the cubic and L-shape rooms, also indicated by the mean RMSE of $0.08$ Pa. Applying domain decomposition and only training and evaluating the receivers in $1/4$ of the domain gives significantly better results with a better fit in both the time and frequency domain, reporting a mean RMSE of $0.03$ Pa on par with the cubic domain.

\subsection{Run-time efficiency}

For real-time applications, we assess the inference time of trained networks with identical layer and neuron configurations, except for the input layer of the branch net, which varies based on the geometry shape. This variation affects parameters and forward propagation performance. Summary information, including parameter count and total storage, is provided in Table \ref{tab:network_sizes}. Using an Nvidia V100 GPU, we predicted impulse responses of length $T=0.5$ s (increased ten times compared to previous experiments), sampled at $f_s=2000$ Hz, for five receiver positions. Execution times for the cubic, L-shape, furnished, and dome geometries were $39$ ms, $49$ ms, $49$ ms, and $132$ ms, respectively. Apart from the dome, these times comfortably meet the real-time threshold of $96$ ms established by previous experiments \citep{sandvad1996dynamic}. Crossing this threshold would introduce significant degradation in azimuth error and elapsed time. The longer execution time for the dome is due to the larger input space covered by discretized input functions, spanning a larger volume compared to other geometries. Our DG-FEM data generation code constructs input function sizes based on the smallest enclosing bounding box and uniform distribution of samples, resulting in unused function values outside the dome geometry. Furthermore, the modified MLP network architecture expands the input layers, increasing the network size compared to a standard MLP network. Convolutional neural networks, as demonstrated in \cite{borrel-jensen_sensitivity_2023}, offer comparable accuracy to modified MLPs while potentially enhancing inference speed.

\subsection{Training time}
Overall the training times for the $3$D geometries are between one and three days on a single GPU. We divide the training time per iteration into data loading and weight/bias update encompassing forward/backward propagation. Our experiments were conducted in the $2$D and $3$D furnished rooms, as summarized in Table S4. Notably, training in $3$D is approximately $64$ times slower per iteration compared to $2$D. In $2$D, the data size of $229$ MB fits in memory, while in $3$D, the data size is $119$ GB, necessitating streaming from disk\footnote{$960$ GB SATA SSD connected to a node at the DTU Computing Center \cite{DTU_DCC_resource}. }. This disparity is the primary reason for the significant increase in training time. Data loading takes $2.1$ seconds in $3$D, while $2$D (loaded from memory) only requires $32.7$ ms. Consequently, the loading time is more than $1\small{,}200\times$ longer in the $3$D scenario. Additionally, the time for weights and biases update is $18\times$ longer in $3$D than in $2$D due to the larger network size, while assuming similar accuracy, as both models in $2$D and $3$D exhibit a mean RMSE of $0.09$ Pa.

\subsection{Transfer learning}
\begin{figure}
\centering
\includegraphics[width=1.0\linewidth]{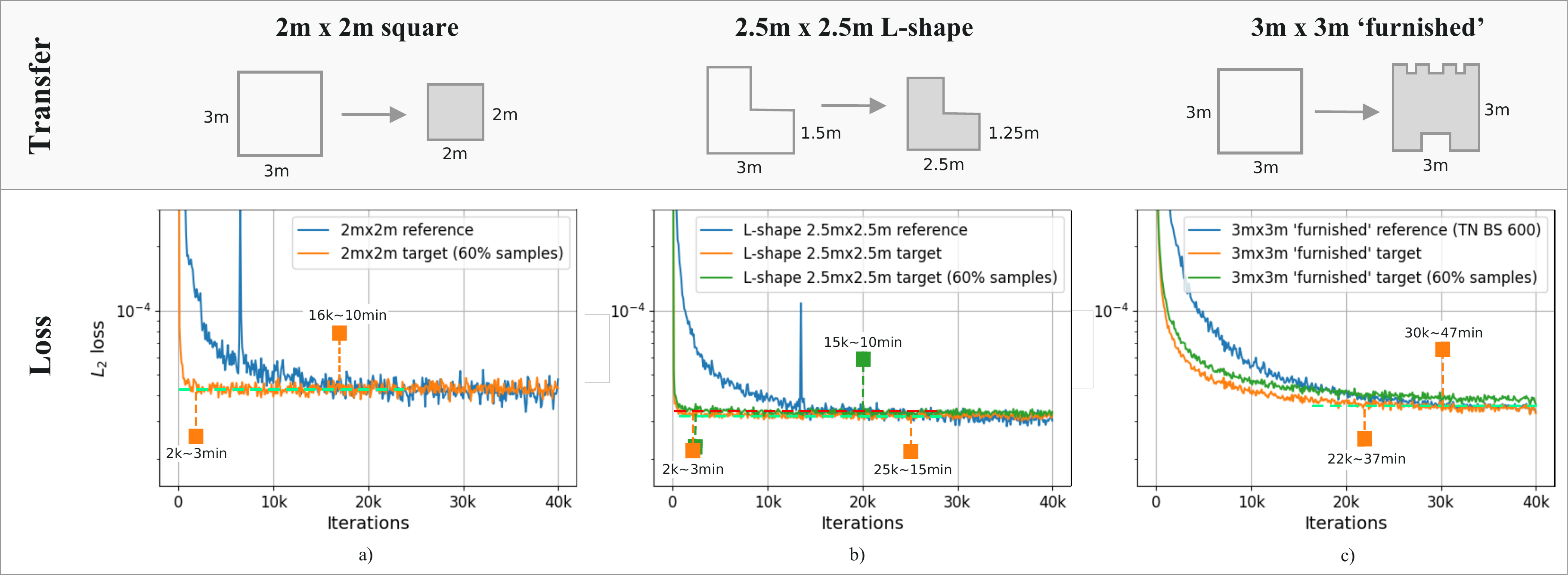}
\caption{Comparative convergence plots of the reference model and the fine-tuned target model for transfer learning scenarios.} \label{fig:transfer_learning}
\end{figure}

In \autoref{fig:transfer_learning}, the convergence rates for training a reference model from scratch and employing a well-trained source model to initialize the weights on a target model followed by fine-tuning are compared. Three cases are considered, $a)$ a square $3 \times 3\text{ m}^2$ to a square $2 \times 2\text{ m}^2$, $b)$ a square $3 \times 3\text{ m}^2$ to a furnished square $3 \times 3\text{ m}^2$, and $c)$ an L-shape geometry with outer dimensions $3 \times 3\text{ m}^2$ to an L-shape geometry with outer dimensions $2.5 \times 2.5\text{ m}^2$. Significant improvements in training time are seen for cases $a)$ and $b)$, with a $3\times$ speedup using only $60\%$ of the data samples on the target domain. Mini-batching for the target model in transfer learning using the spatiotemporal batch size $Q=600$ instead of $Q=200$ for the source model nearly triples the training time but enhances the initial convergence rate, leading to sharper convergence. This effect could also be present in training the reference model, reaching the cross-over point sooner. However, the training time would increase beyond the time saved by a sooner cross-over, making this approach less effective. For example, the L-shape reference would cross at 16k iterations, taking $25$ minutes using $Q=600$, compared to $25$k iterations, taking only $15$ minutes using $Q=200$. The convergence for training the $2 \times 2\text{ m}^2$ rectangular reference model would remain unchanged. In the case of the furnished shape, using the larger batch size narrows the performance gap between the reference and transfer learning in terms of both loss and time. Therefore, the reference model with the larger batch size is chosen for a fair comparison. Additionally, when utilizing only $60\%$ of the samples, the convergence points for the reference and transfer models align earlier for the L-shape and furnished geometries.
\section{Discussion}
The results demonstrate good agreement between the prediction and reference for all five source/receiver pair positions, with RMSE values below $0.10$ Pa. However, some inaccuracies are observed for sharp peaks and high-frequency content in both time and frequency domains, particularly in the furnished room and, to some extent, the full dome, primarily due to the large source position volume and domain volume, respectively. The cubic room shows the best result with a mean RMSE of $0.03$ Pa, compared to $0.09$ Pa for the furnished room, mainly due to the relatively small volume. The dome's source position subdomain has $1\small{,}849$ source positions compared to $2\small{,}826-4\small{,}799$ source positions in the other geometries, which is considerably smaller. However, the predictions are less accurate primarily due to the large volume and, to a lesser extent, due to the geometrical complexity. To investigate the scalability of the DeepONet, instead of increasing the network sizes not necessarily yielding better accuracy/convergence, we applied a simple domain decomposition technique limiting the operator to be evaluated in $1/4$ of the domain, still predicting for all $1\small{,}849$ source positions. This technique showed much-improved accuracy on par with the cubic room, proposing a way to scale these methods to large-scale simulations. Applying the same technique to the furnished room should also increase its accuracy. However, it comes with the disadvantage of training additional DeepONets scaling with the number of partitions.

DeepONet more easily learns lower-frequency modes than higher-frequency modes. Using the \texttt{sine} activation function and employing Fourier expansion in the input layer to span multiple periods accomplishes the goal of helping the network learn higher-frequency content. Although the above modifications dramatically improved the learning capability of DeepONet, it still lacks some accuracy above $700$ Hz for larger and more complex geometries. Increasing the number of layers and neurons did not improve the accuracy, indicating that the network bottleneck is not the capacity but rather difficulties in the optimization to find better optima. It is well-known that all neural networks have challenges when the ratio between upper-frequency and domain size gets larger, which is addressed in the current study by proposing a simple domain decomposition technique.

A spurious noise is observed in the impulse responses before the direct sound arrival, oscillating at the non-causal fundamental frequency. The network struggles to simultaneously learn wave propagation and zero pressures due to the trigonometric feature expansion and \texttt{sine} activation functions. This trade-off between aiding network learning with prior knowledge and learning zero pressures can be managed since the non-zero pressures are small and may not be audible in practical applications.

Training the $2$D domain is efficient, taking less than 40 minutes, while training the $3$D domains requires between one and three days, mainly due to streaming data from disk. This accounts for an overall $64\times$ increase in time, with more than a $1\small{,}200\times$ increase in data loading time for the furnished geometry. The larger network and batch sizes in $3$D contribute to a $18\times$ increase in training time, but the forward/backward propagation time scales better than the cubic complexity of standard numerical methods. Transfer learning experiments show a $3\times$ speedup when fine-tuning the network parameters on scaled geometries for a square and an L-shape domain. Still, limited improvements are seen when transferring to a furnished domain. The results indicate that transfer learning frameworks could lead to faster training, provided that the source and target models are  \textit{similar enough} for the target wave field to be learned efficiently.

The surrogate models exhibit efficient execution, with inference times below $49$ ms for the cubic, L-shape, and furnished rooms, meeting real-time requirements for audio-visual applications. The dome's inference time is slower, taking $132$ ms due to the larger dimensionality of the discretized source input functions. This could be addressed by sampling the input function more accurately, ensuring no zero-samples are outside the geometry. However, if the learned model is intended to be used as a source model for transfer learning, more sophisticated methods should be applied to relate spatial locations between models. Also, using convolutional neural networks for the branch net could decrease the number of network parameters.

To the authors' knowledge, this is the first time a surrogate model with parameterized source positions has been proposed for modeling wave propagation in $3$D domains with realistic frequency-independent and dependent boundaries capable of executing in real-time. These findings are promising, with the potential to overcome current numerical methods' limitations in modeling flexible scenes, such as moving sources. However, further research is needed to address limitations related to larger rooms and better learning of high-frequency content when numerous degrees of freedom are required for source positions. Perceptual studies are also necessary to assess the tolerability of error levels for specific applications.

\section{Governing equations}\label{sec:governing_eq}
The acoustic wave equation for which a surrogate model is to be learned is given as
\begin{equation}
    \frac{\partial^2 p(x,t)}{\partial t^2}-c^2\frac{\partial^2 p(x,t)}{\partial x^2}=0,\qquad t \in \mathbb{R^+}, \qquad x \in \mathbb{R}, \label{wave_equation}
  \end{equation}
  where $p$ is the pressure (Pa), $t$ is the time (s) and $c$ is the speed of sound in air (m/s). The initial conditions (ICs) are satisfied by using a Gaussian impulse function (GIF) as sound source for the pressure part and setting the velocity equal to zero as
  \begin{equation}
  p(x,t=0,x_0) = \exp\left[-\left(\frac{x-x_0}{\sigma_0}\right)^2\right], ~ \frac{\partial p(x,t=0,x_0)}{\partial t} = 0, \label{eq:initial_cond}
  \end{equation}
  with $\sigma_0$ being the width parameter of the pulse determining the frequencies to span (smaller $\sigma_0$ indicates a larger frequency span). The details concerning the boundary condition modeling can be found in \ref{sec:Impedance_boundaries}.

\subsection{Code setup}
JAX 0.4.10 \cite{jax2018}, Flax 0.6.10 \cite{flax2020} and Python 3.10.7 have been used for all experiments and the code is available here: \hyperlink{https://github.com/dtu-act/deeponet-acoustic-wave-prop}{https://github.com/dtu-act/deeponet-acoustic-wave-prop}.

\subsubsection{Data}
The physical speed of sound is $c_\text{phys} = 343$ m/s, and air density is $\rho_0 = 1.2 \text{ kg}/\text{m}^3$, where the speed of sound has been normalized to $c = 1$ m/s to ensure the same resolution in the spatial and temporal dimensions. This is crucial for the optimizer performing gradient descent to find meaningful trajectories. The normalization of the speed of sound can be done trivially by adjusting the time as $t = t_\text{phys}\cdot c_\text{phys}$. Unless stated otherwise, the following will present the results and material parameters in the physical domain. The frequency-independent boundaries are modeled with normalized impedance $\xi_{\text{imp}} = 17.98$, whereas the frequency-dependent boundaries are modeled as a porous material mounted on a rigid backing with thickness $d_\text{mat}=0.03$ m with an airflow resistivity of $\sigma_\text{mat} = 10,000 \text{ Nsm}^{-4}$. This material's surface impedance $Y$ is estimated using Miki's model \cite{Miki199019} and mapped to a six-pole rational function in the form given in \autoref{rational_func} using a vector fitting algorithm \cite{Gustavsen1999} yielding the coefficients for the velocity term from \autoref{velocity_bound_freq_dep}.

The GIF to the branch network has been discretized at the Nyquist limit $\texttt{ppw}=2$. Each sample corresponds to a specific source position, and the number of samples (i.e., the source density) needed for spanning the input space is calculated such that the average resolution between source positions is well-resolved w.r.t. the upper frequency, i.e., $\Delta x_\text{source density} = \frac{c}{f \cdot \text{ ppw}}$.

The $3$D data was generated using a DG-FEM solver \cite{Melander2020}, whereas the $2$D data were generated using an SEM solver \cite{Pind2019}. Ensuring good accuracy at interpolation locations is crucial for the applications of interest. Therefore, the training data was generated with $\texttt{ppw} = 6$ using six-order Jacobi polynomials for all cases except for the dome using fourth-order Jacobi polynomials. The validation and testing data were generated with $\texttt{ppw}=5$ using fourth-order Jacobi polynomials. Hence, we ensure that the mesh points are mostly non-overlapping for the datasets, likewise the Gauss-Lobatto nodes for each element. All simulations span frequencies up to $1\small{,}000$ Hz with an average grid resolution of $\Delta x_{\text{5ppw}} = 0.069$ m and $\Delta x_{\text{6ppw}} = 0.057$ m when using $\texttt{ppw}=5$ and $\texttt{ppw}=6$, respectively. Testing data was generated with five source positions only. 

The time step was $\Delta t = \text{CFL} \times \Delta x / c$ with the Courant-Friedrichs-Lewy constant set to $\text{CFL} = 1.0$ and $\text{CFL} = 0.2$ for frequency-independent and dependent impedance boundaries, respectively. The generated data sets were pruned in the temporal dimensions with $\texttt{ppw}\sim 2$, corresponding to a temporal resolution of $5e^{-4}$ s. Training the models on sparse temporal data results in overfitting, which we exploit for faster training and smaller data sets since interpolation in time is not useful for the applications of interest. The input function $u(x_i)$ for the branch net was uniformly sampled at the Nyquist limit $\texttt{ppw}=2$ in the bounding box enclosing the geometry as depicted in the \autoref{fig:input_function_transfer2D}. This approach facilitates transfer learning but has the disadvantage of unnecessarily large input sizes for non-rectangular domains. The density of the source positions was determined by distributing the source positions with one-fifth wavelength for the training data and roughly one full wavelength for the validation data.

Before training, the spatial data has been normalized as a pre-processing step in the range $[-1,1]$. The temporal dimension is normalized with the spatial normalization factor to ensure equal numerical resolutions in all dimensions of the temporal-spatial domain: e.g., if the spatial data is in the range $\xi \in [-2,2]$ m and the temporal data is in the range $t = [0,10]$ s, then the normalization factor is 2, and the temporal data would be normalized as $t_{\text{norm}} = [0,5]$ s. To summarize, the data set has been constructed as
\begin{align}
    \begin{split}
     \mathcal{D}_j = \{\mathbf{u}_j,\mathbf{\xi}_{i}\}_{i=1}^{N_{\text{full}}}, \quad \text{for} \quad j=1,2\ldots M_{\text{full}}, \quad \text{where} \\
    \mathbf{u}_j = \{u_{j,i}\}_{i=1}^{m}, \qquad \mathbf{\xi}_{i} = \{x_i,y_i,z_i,t_i\},
    \end{split}
\end{align}

$N_{\text{full}}$ is the number of spatiotemporal samples and $M_{\text{full}}$ is the number of (Gaussian) source functions $\mathbf{u}_j$. $\mathbf{u}_j$ is sampled at $m$ fixed sensor locations used as input to the branch net, and $\mathbf{\xi}$ are the spatiotemporal samples used as input to the trunk net.

For training, a single mini-batch for each iteration is compiled by randomly sampling $N$ input sample functions $\{\mathbf{u}^{(i)}\}_{i=1}^{N}$ for the branch net and randomly sampling $Q$ coordinate pairs $\{ \mathbf{\xi}^{(i)} = (x_i,y_i,z_i,t_i) \}_{i=1}^{Q} \in \mathbb{R}^D$ for the trunk net for {\it each} input function. The details of network architecture and mini-batches are provided in \ref{sec:neural_operators}. 

\subsubsection{DeepONet}
The DeepONet architecture used in this work is depicted in \autoref{fig:deeponet_architecture}. In the literature, the DeepONet models have mostly been trained using Gaussian random fields (GRFs) as input to the branch net. However, this work uses the GIF from \autoref{eq:initial_cond} with $\sigma_0 = \frac{c}{\pi\cdot f_\text{max}/2} =  0.22$ m spanning frequencies up to $f_\text{phys} = 1,000$ Hz and is used as a sound source input (initial condition) to the branch net. Using GIFs as ICs drastically reduces the number of samples needed for training compared to GRFs. Limiting the input space to Gaussian functions has no practical limitations in room acoustics since the room impulse response emitting from a GIF can be convolved with any band-limited signal to achieve the acoustical room signal for a fixed frequency range. 

The input to the trunk network is the location $\xi$ where the operator is evaluated and consists of the spatial and temporal coordinates $x,y,z$ and $t$. To overcome the spectral bias \cite{Rahaman2018, basri2020}, the temporal and spatial inputs are passed through a positional encoding mapping as shown in \autoref{eq:fourier_mapping} to learn the high-frequency modes of the data, where the frequencies $\mathbf{f}_{j} = [500,250,167]$ Hz have been chosen relative to the fundamental frequency $f_0 = 1\small{,}000$ Hz, resulting in $2 \times 4 \times 3 = 24$ (\texttt{sine} and \texttt{cos} $\Rightarrow 2$, $x,y,z,t \Rightarrow 4$, expansion terms $\Rightarrow m=3$) additional inputs to the trunk net.
\begin{align}\label{eq:fourier_mapping}
\begin{split}
    \gamma(\mathbf{x}) = \left[\ldots,\cos\left(2\pi f_j \mathbf{x}\right), \sin\left(2\pi f_j \mathbf{x}\right), \ldots\right]^T, \\
    \text{for } j=0,\ldots,m-1.
\end{split}
\end{align}

The modified MLP architecture described in \ref{subsec:deeponet_arch} was used for the branch and trunk net. Self-adaptive weights were applied to all spatiotemporal locations optimized using a separate ADAM optimizer. Gradient clipping with an absolute value of $0.1$ was needed to limit fluctuations that could sometimes make the optimizer jump to a drastically larger loss. The weights of the networks are initialized \cite{Sitzmann2020} as $w_i \thicksim \mathcal{U}\left(-\frac{\sqrt{6/n}}{k}, \frac{\sqrt{6/n}}{k}\right)$, where $n$ denotes the number of input neurons to the $i$'th neuron and $k$ was empirically chosen as $k=30$ for all layers except for $k=1$ used at the first layer. The first layer is initialized with weights such that the sine functions $\sin(w_0\cdot \mathbf{Wx}+\mathbf{b})$ spans multiple periods, where the angular frequency $w_0 = 30$ was empirically found to give the best results.

\subsubsection{Domain decomposition}
When the frequency range is increased or, correspondingly, the domain size is increased, the accuracy of the deep neural network will decrease for a fixed network size. Increasing the network size in terms of layers and neurons should theoretically be sufficient to regain the required accuracy; however, this is often not the case. This is well-known in the literature and applies to, but is not limited to, both DeepONet and PINNs. Domain decomposition approaches, such as XPINNs \cite{jagtap2020extended} applied for PINNs, have been shown to overcome these limitations, but with the expense of more neural networks to train. The general idea is to split the domain into (non-overlapping) partitions, each running separate neural networks and adding an additional loss term at the interface, imposing continuity conditions. In this work, training the DeepONet is purely data-driven, and a simpler approach has been taken. We divide the full domain $\Omega \in \mathbb{R}^3$ into four non-overlapping partitions $\Omega = \Omega_1 \cup \Omega_2 \cup \Omega_3 \cup \Omega_4$, 
where $\{x_i^{(k)},y_i^{(k)},z_i^{(k)}\}_{i=1}^{N_k} \in \Omega_k$, $k=1\ldots4$ and $N = \sum_{k=1}^4 N_i$, $N_i \in \mathbb{Z}^+$. We then train four DeepONets $\mathcal{NN}_k$, each on the full source function space $\mathbf{u}$, but restrict the location where we evaluate the operator at one of the partitions $\Omega_i$. The temporal samples are kept (could also be partitioned if needed), which gives us the data set $\mathcal{D}$ for training a DeepONet $\mathcal{NN}_k$ for a $k$'th partition
\begin{eqnarray}
\begin{split}
    \mathcal{D}^{(k)}_j = \{\mathbf{u}_j,\mathbf{\xi}^{(k)}_{i}\}_{i=1}^{N_k}, \quad \text{for} \quad j=1,2\ldots N, \quad \text{where} \\
    \mathbf{\xi}_i^{(k)} = \{x_i^{(k)},y_i^{(k)},z_i^{(k)},t_i\}.
\end{split}
\end{eqnarray}
This work does not enforce continuity at the interfaces but could be done by calculating the mean of overlapping domains near the interfaces.

\subsubsection{Self-adaptive weights}
Weighting individual samples in the loss function can be advantageous for the DeepONet to perform better. Using point-wise weights, the loss function can be minimized w.r.t. the network parameters but maximized w.r.t. the point-wise loss weights. This approach, called self-adaptive weights, was originally introduced to improve the performance for PINNs \cite{mcclenny_self-adaptive_2023} and later extended to DeepONet \cite{kontolati2022influence}. The self-adaptive weights are applied to all spatiotemporal samples and initialized to 1. To ensure stability in case some sample points are not converging, the weights have been clamped to take values between $0$ and $1\small{,}000$. A separate ADAM optimizer was used for updating the self-adaptive weights with a learning rate two orders of magnitude lower than the learning rate for the network parameters.
 
\subsubsection{Transfer learning}
Training DeepONet surrogate models for every geometry might get intractable for real-world usage due to the time and resources needed to train realistic $3$D geometries. A more tractable strategy that could be applied for real-world problems is to pre-train DeepONets for geometries with certain traits (e.g., cubic rooms, L-shaped rooms, penta shapes, furnished rooms, etc.) and fine-tune the training on specific target room geometry by transferring the weight from a pre-trained DeepONet corresponding to the closest-matching geometry. We have made an investigation in $2$D by performing transfer learning between rectangular, L-shape, and furnished geometries of varying sizes. First, the source models are trained using a network with two hidden layers of width $2\small{,}048$ for both the branch and the trunk net using mini-batching of $N=64$ and $Q=\{200,600\}$. Then, the optimized network parameters are used to initialize the target model, a subset of the layers are frozen, and the new model is fine-tuned on data corresponding to the new geometry with $N=64$ and $Q=\{200,600\}$ on the full training set or a subset using only $60 \%$ of the Gaussian input functions (i.e., source positions). When freezing layers, the optimizer will skip updating the corresponding weights and biases for these. By sampling the Gaussian input function on an enclosing rectangle, the mapping from the target model's source positions to the source model's closest corresponding source positions can be done straightforwardly as shown in the \autoref{fig:input_function_transfer2D}. It is also important to ensure that the spatial locations between the source and target model are as closely related as possible. For all the cases, the spatial alignment between source and receiver is done at the coordinate $[0,0]$ m. The first hidden layer is frozen in the trunk net, leaving the second (non-linear) hidden layer and the linear output layer trainable. In contrast, only the linear output layer is trainable in the branch net. From the experiments, the trunk net learning the basis function is more important to fine-tuning the new geometry than the basis function coefficients learned by the branch net.

\section{Acknowledgements}
This research was conducted using computing resources and services at the Center for Computation and Visualization, Brown University, and at DTU Computing Center \cite{DTU_DCC_resource}. A big thanks go to Rômulo Silva for fruitful discussions.
SG and GEK would like to acknowledge support by the MURI-AFOSR FA9550-20-1-0358 project.

\bibliographystyle{elsarticle-num} 
\bibliography{deeponet}
\newpage
\appendix
\section{Parameterized PDEs in acoustics}
The challenge of utilizing parametric PDEs has motivated increased research. Reduced order methods \cite{Hesthaven2015, Llopis2022} aim to reduce the degrees of freedom; however, despite achieving orders of magnitude in accelerations for many applications, these techniques still cannot meet the runtime requirement for real-time experiences for sound propagation in realistic 3D schenes. Recently, the possibility of generating surrogate models with little data was demonstrated using physics-informed neural networks \cite{Raissi2019} and applied for acoustics problems in \cite{nborrel2021}. Previous attempts to overcome the storage requirements of the IR include work for lossy compression \cite{Raghuvanshi2014}. Lately, a novel portal search method has been proposed as a drop-in solution to pre-computed IRs to adapt to flexible scenes, e.g., when doors and windows are opened and closed \cite{raghuvanshi2021dynamic}.

\section{Methods}
\subsection{Neural operators}\label{sec:neural_operators}
Let $\Omega \subset \mathbb{R}^D$ be a bounded open set and $\mathcal{U}=\mathcal{U}(\Omega; \mathbb{R}^{d_x})$ and $\mathcal{Y}=\mathcal{Y}(\Omega;\mathbb{R}^{d_y})$ two separable Banach spaces. Furthermore, assume that $\mathcal{G}: \mathcal{U} \rightarrow \mathcal{Y}$ is a non-linear map arising from the solution of a time-dependent PDE. The objective is to approximate the nonlinear operator via the following parametric mapping
\begin{equation}
\begin{aligned}
    \mathcal{G}: \mathcal{U} \times \Theta \rightarrow \mathcal{Y} \hspace{15pt} \text{or}, \hspace{15pt} \mathcal{G}_{\theta}: \mathcal{U} \rightarrow \mathcal{Y}, \hspace{5pt} \theta \in \Theta
\end{aligned}
\end{equation}
where $\Theta$ is a finite-dimensional parameter space. The optimal parameters $\theta^*$ are learned via the training of a neural operator with backpropagation based on a dataset $\{\mathbf{u}_j, \mathbf{y}_j \}_{j=1}^N$ generated on a discretized domain $\Omega_m = \{x_1, \dots, x_m\} \subset \Omega$ where $\{x_j\}_{j=1}^m$ represent the sensor locations, thus $\mathbf{u}_{j|\Omega_m} \in \mathbb{R}^{D_x}$ and $\mathbf{y}_{j|\Omega_m} \in \mathbb{R}^{D_y}$ where $D_x= d_x \times m$ and $D_y = d_y \times m$. 

\subsubsection{The deep operator network (DeepONet)}
\begin{figure}
\centering
\includegraphics[width=0.8\linewidth]{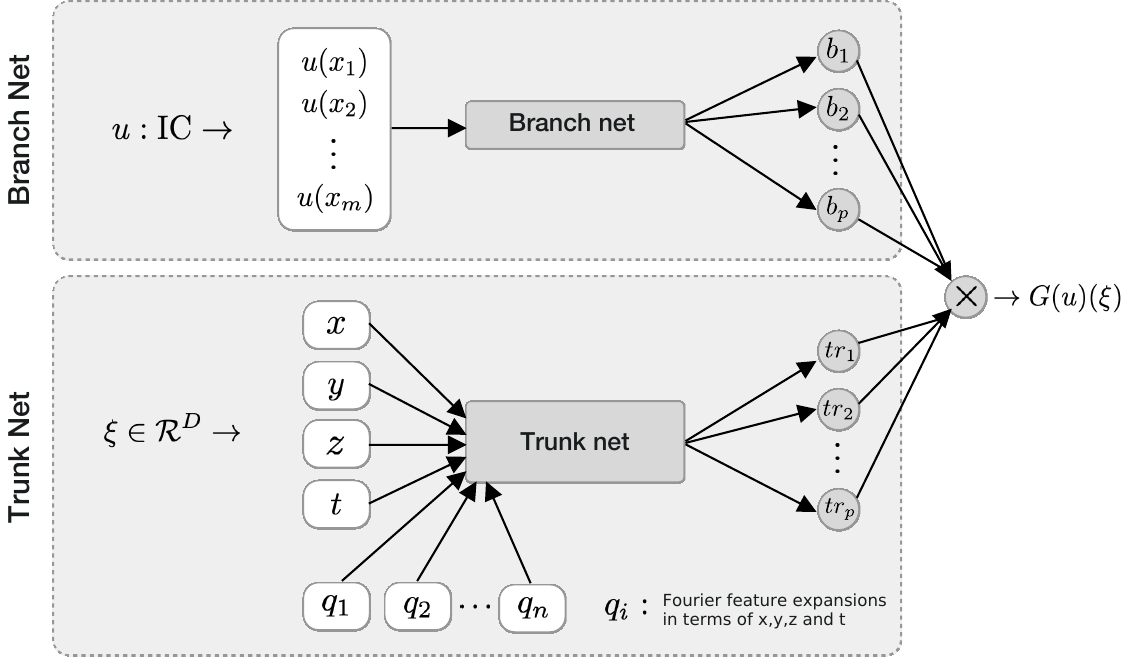}
\caption{DeepONet architecture with parameterized source position for predicting the impulse response for a source/receiver pair over time for a 3D domain. The branch net is taking as input a Gaussian source function $\mathbf{u}$ determining the source position, sampled at fixed sensor locations. The spatial coordinates $x$, $y$, $z$, and temporal coordinate $t$ are denoted by $\xi$ and are used as input to the trunk net mapping into the output domain of the operator. }
\label{fig:deeponet_architecture}

\end{figure}

DeepONet \cite{Lu2021} aims to learn operators between infinite-dimensional Banach spaces. Learning is performed in a general setting in the sense that the sensor locations $\{x_i\}_{i=1}^m$ at which the input functions are evaluated need not be equispaced; however, they need to be consistent across all input function evaluations. Instead of blindly concatenating the input data (input functions $[\mathbf{u}(x_1), \mathbf{u}(x_2), \dots, \mathbf{u}(x_m)]^T$ and locations $\zeta$)
as one input, i.e., $[\mathbf{u}(x_1), \mathbf{u}(x_2), \dots, \mathbf{u}(x_m), \zeta]^T$, DeepONet employs two subnetworks and treats the two inputs equally. Thus, DeepONet can be applied for high-dimensional problems where the dimension of $\mathbf{u}(u_i)$ and $\zeta$ no longer match since the latter is a vector of $d$ components in total. A trunk network $\mathbf{f}(\cdot)$, takes as input $\zeta$ and outputs $[tr_1, tr_2, \ldots, tr_p]^T \in \mathbb{R}^p$ while a second network, the branch net $\mathbf{g}(\cdot)$, takes as input $[\mathbf{u}(x_1), \mathbf{u}(x_2), \dots, \mathbf{u}(x_m)]^T$ and outputs $[b_1, b_2, \ldots, b_p]^T \in \mathbb{R}^p$. Both subnetwork outputs are merged through a dot product to generate the quantity of interest. A bias $b_0 \in \mathbb{R}$ is added in the last stage to increase expressivity, i.e., $\mathcal{G}(\mathbf{u})(\zeta) \approx \sum_{i=k}^p b_k t_k + b_0$. The generalized universal approximation theorem for operators, inspired by the original theorem introduced by \cite{Chen1995}, is presented below. The generalized theorem essentially replaces shallow networks used for the branch and trunk net in the original work with deep neural networks to gain expressivity. An overview of the architecture used in this work is depicted in \autoref{fig:deeponet_architecture}.

\begin{theorem}[Generalized Universal Approximation Theorem for Operators.]
\label{pythagorean}
Suppose that $X$ is a Banach space, $K_1 \subset X$, $K_2 \subset \mathbb{R}^d$ are two compact sets in $X$ and $\mathbb{R}^d$, respectively, $V$ is a compact set in $C(K_1)$. Assume that: $\mathcal{G}: V \rightarrow C(K_2)$ is a nonlinear continuous operator. Then, for any $\epsilon > 0$, there exist positive integers $m, p$, continuous vector functions $\mathbf{g}: \mathbb{R}^m \rightarrow \mathbb{R}^p$, $\mathbf{f}: \mathbb{R}^d \rightarrow \mathbb{R}^p$, and $x_1, x_2, \dots , x_m \in K_1$ such that   
\[  \Bigg\lvert \mathcal{G}(\mathbf{u})(\zeta) - \langle  \underbrace{\mathbf{g}(\mathbf{u}(x_1), \mathbf{u}(x_2), \ldots, \mathbf {x}(x_m))}_{\text{branch}}, \underbrace{\mathbf{f}(\zeta)}_{\text{trunk}}  \rangle \Bigg\rvert < \epsilon \]
holds for all $\mathbf{u} \in V$ and $\zeta \in K_2$, where $\langle \cdot, \cdot \rangle$ denotes the dot product in $\mathbb{R}^p$. For the two functions $\mathbf{g}, \mathbf{f}$ classical deep neural network models and architectures can be chosen that satisfy the universal approximation theorems of functions, such as fully-connected networks or convolutional neural networks.
\end{theorem}

The method accurately learns the mapping from an input space of functions into a space of output functions, thereby generalizing the solution for a parametrized PDE. DeepONet provides a simple architecture that is fast to train, utilizing data from high-fidelity simulations describing sound propagation, and allows for continuous target outputs predicting source/receiver pairs in a grid-less domain almost instantly. 

The Fourier neural operator \cite{Li2020}, Wavelet neural operator \cite{tripura2023wavelet}, and the Laplace neural operator \cite{cao2023lno} are a separate class of neural operator where the solution operator is expressed as an integral operator of Green's function that is parameterized in the Fourier, Wavelet, and Laplace space, respectively. All these versions are different realizations of DeepONet if appropriate changes are imposed on its architecture. Approximating operators is a paradigm shift from current and established machine learning techniques focusing on function approximation to the solution of the PDEs.

\subsubsection{DeepONet architecture}\label{subsec:deeponet_arch}
\begin{figure}
\centering
\includegraphics[width=0.8\textwidth]{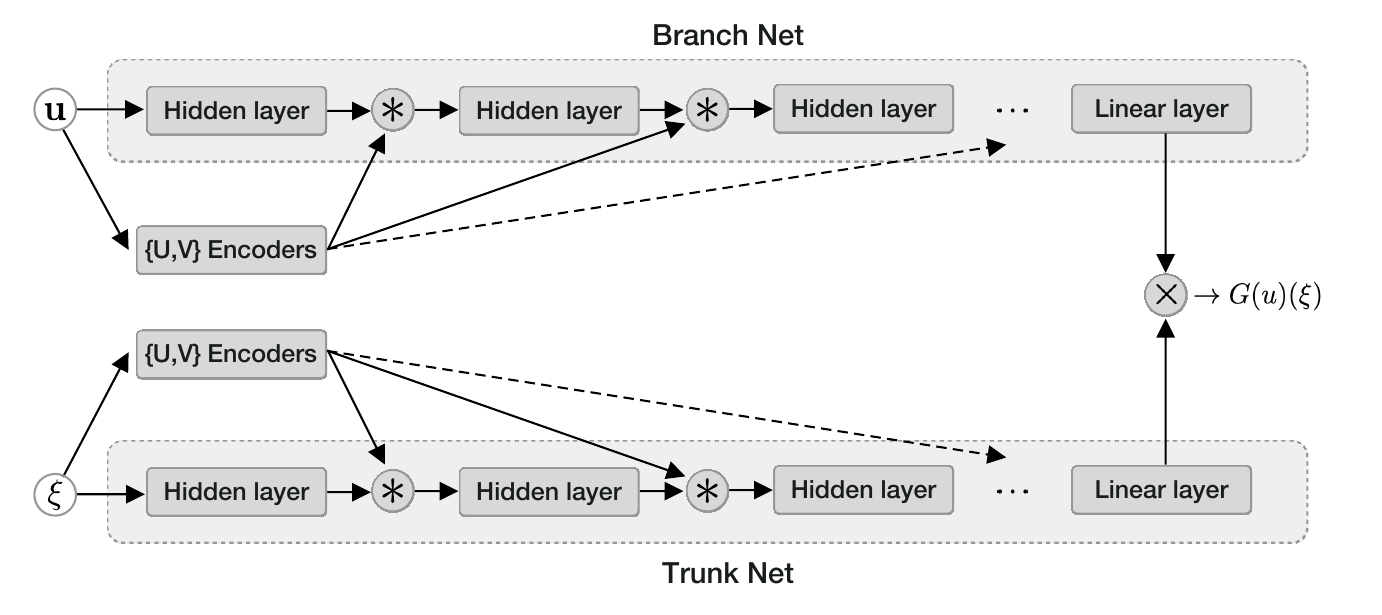}  
\caption{The modified MLP architecture applied for the DeepONet. Two encoders $u$ and $v$ implemented as single-layer neural networks are applied for each MLP, embedding the inputs into a latent space with the size of the layer width of the MLP. The embedded features are then inserted into each hidden layer illustrated by `*' performing the operation $(1 - f_i) \odot u + f_i \odot v$. }
\label{fig:mod_mlp}
\end{figure}

The DeepONet framework allows many network architectures, such as feed-forward neural networks (FNN), multi-layer perception (MLP), recurrent neural networks (RNN), convolutional neural networks (CNN), graph neural networks (GNN), and convolutional graph neural networks (CGNN). In this work, we have used a modification to the MLP (mod-MLP) for both the branch and trunk net originally proposed in \cite{WangParis2021_understanding_gradient} for PINNs and in \cite{WangParis2021} for DeepONets shown to outperform the conventional FNNs. First, let us define a standard FNN consisting of an input layer $\mathbf{x}$, $n$ hidden layers, and an output layer. The mapping from an input $\mathbf{x}$ to an output $\mathbf{y}$ is defined as
\begin{align}
\begin{split}
    \mathbf{y} &= (f_0 \circ f_1 \circ \ldots \circ f_n)(\mathbf{x}), \\
    f_i(\mathbf{x}) &= \sigma_i(\mathbf{W}^{i}\mathbf{x} + \mathbf{b}^i).
\end{split}    
\end{align}
$\sigma_i(\mathbf{x})$ is a non-linear activation function (except for a linear activation in the last layer), where $\mathbf{W}^i$ and $\mathbf{b}^i$ are the weight and bias parameters to learn. An MLP is a special case of an FNN, where every layer is fully connected, and the number of nodes in each layer is the same. The key extension is the introduction of two encoder networks encoding the input variables to a higher-dimensional feature space. The networks consisting of a single layer are shared between all layers, and a pointwise multiplication operation is performed to update the hidden layers. Let the two shallow encoder networks with width size equal to the hidden layers be denoted $u(\mathbf{x})$ and $v(\mathbf{x})$ and defined as a simple perceptron
\begin{align}
\begin{split}
    u(\mathbf{x}) = \sigma(\mathbf{W}_u\mathbf{x} + \mathbf{b}_u), \\
    v(\mathbf{x}) = \sigma(\mathbf{W}_v\mathbf{x} + \mathbf{b}_v),
\end{split}
\end{align}
then the mod-MLP is defined as
\begin{equation}
\begin{split}
\mathbf{y} =& ~ ((1 - f_0) \odot u + f_0 \odot v)~\circ \\
                & ~((1 - f_1) \odot u + f_1 \odot v) ~\circ \\
                & \qquad\qquad\qquad \vdots \\
                & ~((1 - f_n) \odot u + f_n \odot v)(\mathbf{x}),
\end{split}
\end{equation}
where $\odot$ denotes elementwise multiplication, $\circ$ is the function composition operator, and $\mathbf{W}_{\{u,v\}}$ and $\mathbf{b}_{\{u,v\}}$ are the weights and biases for the two encoder networks. The architecture is depicted in \autoref{fig:mod_mlp}, where the encoder networks are applied separately for the branch and trunk net. This implementation differs from the implementation in \cite{wang_improved_2022}, where only two encoder networks are shared between the branch and trunk layer. The motivation behind the mod-MLP is to better propagate information stably through the network since the trainability of the DeepONet depends on merging the branch and trunk net in terms of their inner product only in the last layer. Hence, if the input signals are not properly propagated through the network, this may lead to ineffective training and poor model performance.

\subsubsection{DeepONet setup}
Five hidden layers with $2{\small,}048$ neurons each were used for the branch and trunk net in $3$D; two hidden layers with $2{\small,}048$ neurons each were used for the branch and trunk net in $2$D. The ADAM optimizer and the mean-squared error for calculating the losses were used with a learning rate of $1e-3$ and exponential decay of $0.90$ per $2\small{,}000$ iterations for all experiments. Self-adaptive weights were applied to all spatiotemporal locations using a separate ADAM optimizer with a learning rate two orders of magnitude smaller than the learning rate of the optimizer used for the network parameters. All experiments used mini-batches of $N=64$, $Q=1{\small,}000$, except for the dome, where mini-batches of $N=96$, $Q=1{\small,}500$ were used. For the transfer learning in 2D, $N=64$ and $Q=\{200,600\}$ were used for training the reference and target models. Note, that the data set batch dimensions $\mathbf{u}$, $\mathbf{\xi}$, $G(\mathbf{u})(\mathbf{\xi})$ are $(N \times Q, m)$, $(N \times Q, D)$, $(N \times Q, 1)$, respectively.

\subsection{Impedance boundaries}\label{sec:Impedance_boundaries}
We consider impedance boundaries and denote the boundary domain as $\Gamma$. We will omit the source position $x_0$ in the following. For frequency-independent impedance boundaries, the acoustic properties of a wall can be described by its surface impedance $Z_s = \frac{p}{v_n}$  \cite{kuttruff2016room} where $v_n$ is the normal component of the velocity at the same location on the wall surface. Combining the surface impedance with the pressure term $\frac{\partial p}{\partial \mathbf{n}} = -\rho_0 \frac{\partial v_n}{\partial t}$ of the linear coupled wave equation yields
\begin{equation}
	\frac{\partial p}{\partial t} = -c\xi_{\text{imp}}\frac{\partial p}{\partial \mathbf{n}}, \quad \text{for} \quad \Omega  \quad \text{and} \quad t\geq0, \label{eq:impedance_bound_cond}
\end{equation}
where $\xi_{\text{imp}} = Z_s/(\rho_0c)$ is the normalized surface impedance and $\rho_0$ denotes the air density ($\text{kg}/\text{m}^3$). Note that perfectly reflecting boundaries can be obtained by letting $\xi_{\text{imp}} \rightarrow \infty$ be the Neumann boundary formulation.

For frequency-dependent impedance boundaries, the wall impedance can be written as a rational function in terms of the admittance $Y = 1/Z_s$ and rewritten by using partial fraction decomposition in the last equation \cite{Troian2017}
\begin{align}
\begin{split}
  Y(\omega) &= \frac{a_0 + \ldots + a_N(-i\omega)^N}{1+\ldots+b_N(-i\omega)^N} \\
    &= Y_{\infty} + \sum_{k=0}^{Q-1}\frac{A_k}{\lambda_k - i\omega} + \sum_{k=0}^{S-1}\left( \frac{B_k + iC_k}{\alpha_k + i\beta_k - i\omega} + \frac{B_k - iC_k}{\alpha_k -i\beta_k -i\omega} \right),
\end{split} \label{rational_func}
\end{align}
where $a,b$ are real coefficients, $i = \sqrt{-1}$ being the complex number, $Q$ is the number of real poles $\lambda_k$, $S$ is the number of complex conjugate pole pairs $\alpha_k \pm j\beta_k$, and $Y_{\infty}$, $A_k$, $B_k$ and $C_k$ are numerical coefficients. Since we are concerned with the (time-domain) wave equation, the inverse Fourier transform is applied to the admittance and the partial fraction decomposition term in \autoref{rational_func}. Combining these gives \cite{Troian2017}
\begin{align}
  v_n(t) = Y_{\infty}p(t) + \sum_{k=0}^{Q-1}A_k\phi_k(t) + \sum_{k=0}^{S-1}2\left[ B_k \psi_k^{(0)}(t) + C_k\psi_k^{(1)}(t) \right]. \label{velocity_bound_freq_dep}
\end{align}
The functions $\phi_k$, $\psi_k^{(0)}$, and $\psi_k^{(1)}$ are the so-called accumulators determined by the following set of ordinary differential equations (ODEs) referred to as auxiliary differential equations (ADEs)
\begin{equation}
\begin{aligned}
  \frac{d\phi_k}{dt} + \lambda_k\phi_k = p, \qquad
  \frac{d\psi_k^{(0)}}{dt} + \alpha_k\psi_k^{(0)} + \beta_k\psi_k^{(1)} = p, \qquad
  \frac{d\psi_k^{(1)}}{dt} + \alpha_k\psi_k^{(1)} - \beta_k\psi_k^{(0)} = 0. \label{ade_diff_eq}
\end{aligned}
\end{equation}
The boundary conditions can then be formulated by inserting the velocity $v_n$ calculated in \autoref{velocity_bound_freq_dep} into the pressure term of the linear coupled wave equation $\frac{\partial p}{\partial \mathbf{n}} = -\rho_0 \frac{\partial v_n}{\partial t}$. 

\begin{table}
    \centering    
    \begin{tabular}{c|cc|ccc}
         & \multicolumn{2}{c|}{Source function} & \multicolumn{3}{c}{Location} \\
         & Count & Sensors & Time steps & Mesh points & Total per source \\
         \hline\hline
         T - Cubic & $2\small{,}826$ & $28$ & $101$ & $57\small{,}124$ & $5.8$M \\
         V - Cubic & $100 $  & $28$ & $101$ & $17\small{,}643$ & $1.8$M \\
         \hline\hline
         T - L-shape & $5\small{,}165$ & $3\small{,}888$ & $101$ & $93\small{,}675$ & $9.5$M \\
         V - L-shape & $180$   & $3\small{,}888$ & $101$ & $51\small{,}201$ & $5.2$M \\
         \hline\hline
         T - Furn. & $4\small{,}799$ & $3\small{,}888$ & $101$ & $123\small{,}994$ & $12.5$M \\
         V - Furn. & $203$   & $3\small{,}888$ & $101$ & $74\small{,}819$ & $7.6$M \\
         \hline\hline         
         T - Dome & $1\small{,}849$ & $19\small{,}602 $& $101$ & $213\small{,}130$ & $21.5$M \\
         V - Dome & $94$    & $19\small{,}602$ & $101$ & $165\small{,}025$ & $16.7$M* \\
         \hline\hline
         T - Dome $1/4$ & $1\small{,}849$ & $19\small{,}602$ & $101$ & $51,665$ & $5.2$M \\
         V - Dome $1/4$ & $94$    & $19\small{,}602$ & $101$ & $40,240$ & $4.1$M* \\
         \hline\hline
    \end{tabular}
    \caption{Data sizes for the four geometries. The data has been saved in 16-bit floating point precision. The dome $1/4$ arises from being spatially partitioned into four partitions, subsequently evaluated at one partition only. `T' denotes training data, `V' denotes validation data. *Note that the mesh point ratio between training and validation data differs for the dome compared to the other geometries. This is caused by the meshing algorithm forcing finer resolutions in regions near the sphere to capture the complex geometry. }
    \label{tab:dataset_sizes}
\end{table}

\begin{table}
\begin{center}    
    \begin{tabular}{c||c|c|c|c|c||c}
    \multicolumn{1}{c||}{} & \multicolumn{1}{c|}{$s_1$} & \multicolumn{1}{c|}{$s_2$} & \multicolumn{1}{c|}{$s_3$} & \multicolumn{1}{c|}{$s_4$} & \multicolumn{1}{c||}{$s_5$} & \multicolumn{1}{c}{Mean}\\
    Domain & RMSE & RMSE & RMSE & RMSE & RMSE & RMSE \\    
    \hline\hline
    Cubic & $0.03$ Pa & $0.03$ Pa & $0.02$ Pa & $0.04$ Pa & $0.03$ Pa & $0.03$ Pa \\
    \hline
    L-shape & $0.06$ Pa & $0.04$ Pa & $0.05$ Pa & $0.04$ Pa & $0.04$ Pa & $0.05$ Pa \\
    \hline
    Furnished & $0.09$ Pa & $0.09$ Pa & $0.09$ Pa & $0.08$ Pa & $0.08$ Pa & $0.09$ Pa \\
    \hline
    Dome & $0.08$ Pa & $0.05$ Pa & $0.08$ Pa & $0.10$ Pa & $0.10$ Pa & $0.08$ Pa \\
    \hline
    Dome $1/4$ & $0.03$ Pa & $0.02$ Pa & $0.04$ Pa & $0.04$ Pa & $0.04$ Pa & $0.03$ Pa \\
    \hline
    \end{tabular}
    \caption{Impulse receiver errors for source/receiver pairs $s_i$ given in the text. The root mean square error (RMSE) is used to access the errors, defined as $\text{RMSE} = \sqrt{ \frac{\sum_{n=1}^{N}(p_{\text{ref}_i} -  p_{\text{pred}_i})^2}{N} }$.
    }\label{tab:pred_errors}
\end{center}
\end{table}

\begin{table}[]
    \centering    
    \begin{tabular}{c|cc|cc|cc}
         & \multicolumn{2}{c|}{Layer}& inputs & outputs & param. & size \\
         \hline\hline
         \multirow{5}{*}{\rotatebox[origin=c]{90}{\footnotesize Cubic }} &
         \multirow{5}{*}{\rotatebox[origin=c]{90}{\footnotesize Branch net }} &
             U,V               & $1{\small,}728$ & $2{\small,}048$ & $2\times3.5$M & $2\times14$ MB \\
         &&  Hidden layer (in)  & $1{\small,}728$ & $2{\small,}048$ & $3.5$M & $14$ MB \\
         &&  Hidden layers      & $2{\small,}048$ & $2{\small,}048$ & $4\times4.2$M   & $4\times17$ MB \\
         &&  Output layer       & $2{\small,}048$ & $100$  & $204$k     & $820$ KB\\
         \cline{3-7}
         &&  Total              & -     & -   & $27.6$M & $111$ MB \\
         \hline\hline
         \multirow{5}{*}{\rotatebox[origin=c]{90}{\footnotesize Furnished }} &
         \multirow{5}{*}{\rotatebox[origin=c]{90}{\footnotesize Branch net }} &
             U,V               & $3{\small,}888$ & $2{\small,}048$ & $2\times8$M & $2\times32$ MB \\
         &&  Hidden layer (in)  & $3{\small,}888$ & $2{\small,}048$ & $8$M & $32$ MB \\
         &&  Hidden layers      & $2{\small,}048$ & $2{\small,}048$ & $4\times4$M & $4\times17$ MB \\
         &&  Output layer       & $2{\small,}048$ & $100$  & $204$k   & $820$ KB\\
         \cline{3-7}
         &&  Total              & -     & -   & $40.9$M & $164$ MB \\
         \hline\hline
         \multirow{5}{*}{\rotatebox[origin=c]{90}{\footnotesize L-shape }} &
         \multirow{5}{*}{\rotatebox[origin=c]{90}{\footnotesize Branch net }} &
             U,V               & $3{\small,}888$ & $2{\small,}048$ & $2\times8$M & $2\times32$ MB \\
         &&  Hidden layer (in)  & $3{\small,}888$ & $2{\small,}048$ & $8$M & $32$ MB \\
         &&  Hidden layers      & $2{\small,}048$ & $2{\small,}048$ & $4\times4$M & $4\times17$ MB \\
         &&  Output layer       & $2{\small,}048$ & $100$  & $204$k   & $820$ KB\\
         \cline{3-7}
         &&  Total              & -     & -   & 40.9M & $164$ MB \\
         \hline\hline
         \multirow{5}{*}{\rotatebox[origin=c]{90}{\footnotesize Dome }} &
         \multirow{5}{*}{\rotatebox[origin=c]{90}{\footnotesize Branch net }} &
             U,V               & $19{\small,}602$ & $2{\small,}048$ & $2\times40.1$M & 2$\times$161MB \\
         &&  Hidden layer (in)  & $19{\small,}602$ & $2{\small,}048$ & $40.1$M & $161$ MB \\
         &&  Hidden layers      & $2{\small,}048$   & $2{\small,}048$ & $4\times4.2$M  & $4\times17$ MB \\
         &&  Output layer       & $2{\small,}048$   & $100$  & $204$k      & $820$ KB\\
         \cline{3-7}
         &&  Total              & -     & -    & $137$M & $549$ MB \\
         \hline\hline
         \multirow{5}{*}{\rotatebox[origin=c]{90}{\footnotesize [all] }} &
         \multirow{5}{*}{\rotatebox[origin=c]{90}{\footnotesize Trunk net }} &
             U,V               & $28$    & $2{\small,}048$ & $2\times59$k    & $2\times238$k \\
         &&  Hidden layer (in)  & $28$    & $2{\small,}048$ & $59$k    & $238$k \\
         &&  Hidden layers      & $2{\small,}048$  & $2{\small,}048$ & $4\times4$M & $4\times17$ MB \\
         &&  Output layer       & $2{\small,}048$  & $100$  & $204$k   & $820$ KB\\
         \cline{3-7}
         &&  Total              & -     & -    & $17$M & $69$ MB \\
         \hline\hline
    \end{tabular}
    \caption{Branch and trunk network parameters for the cubic, L-shape, furnished, and dome geometries. The input function to the branch net has been uniformly sampled on the enclosed bounding box for the geometries; why the input size is the same for the L-shape and furnished rooms both having outer dimension $3\text{m}\times3\text{m}\times2\text{m}$. The trunk net is the same for all geometries.}
    \label{tab:network_sizes}
\end{table}

\begin{table}
    \centering
    \begin{tabular}{c||ccc|c}
        \multicolumn{1}{c||}{} & \multicolumn{3}{c|}{\footnotesize \bf Timings} & \multicolumn{1}{c}{\footnotesize \bf Data size} \\
         & per iter & loading & back-prop & per iter \\
        \hline\hline
        $2$D Furn. & $32.7\text{ ms}$ & $1.3\text{ ms}/3.8\%$   & $31.4\text{ ms}/96.2\%$  & $0.024$ MB \\
        $3$D Furn. & $2.1$ s  & $1.6\text{ s}/73\%$ & $564\text{ ms}/27\%$ & $1{\small.}5$ GB \\
        \hline
        Factor $3$D/$2$D & 64$\times$ & 1230$\times$ & 18$\times$ & 62,500$\times$ \\
        \hline
    \end{tabular}
    \caption{Training time divided into data loading and weight/bias updates through forward/back-propagation. The timings are given per iteration step for the furnished room in 2D and 3D. The 2D network has two layers of width $2\small{,}048$ for the BN and TN with batch size $64 \times 200 = 12\small{,}800$, and all data fits into memory for fast access and efficient sampling. The 3D network has five layers of width $2\small{,}048$ for the BN and TN with batch size $64 \times 1\small{,}000 = 64\small{,}000$. The data is stored in HDF5 format in separate files for each source position. Therefore, the source position can be sampled randomly by loading a subset of the HDF5 files. In contrast, the temporal/spatial data cannot efficiently be accessed randomly on disk. Therefore all data for each file are loaded in memory, taking up $64 \times 101 \times 123\small{,}994$ 16-bit samples (source sample $\times$ temporal dim. $\times$ spatial dim).} \label{tab:timings} 
\end{table}

\begin{figure}
\centering
\includegraphics[width=0.8\linewidth]{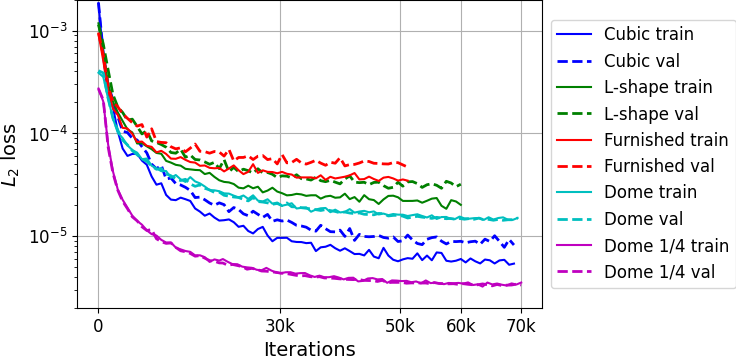}
\caption{Convergence plot showing the training and validation loss for the cubic, L-shape, furnished, and dome geometries. }
\label{fig:convergence_plot}
\end{figure}

\begin{figure}
\centering
\includegraphics[width=0.7\linewidth]{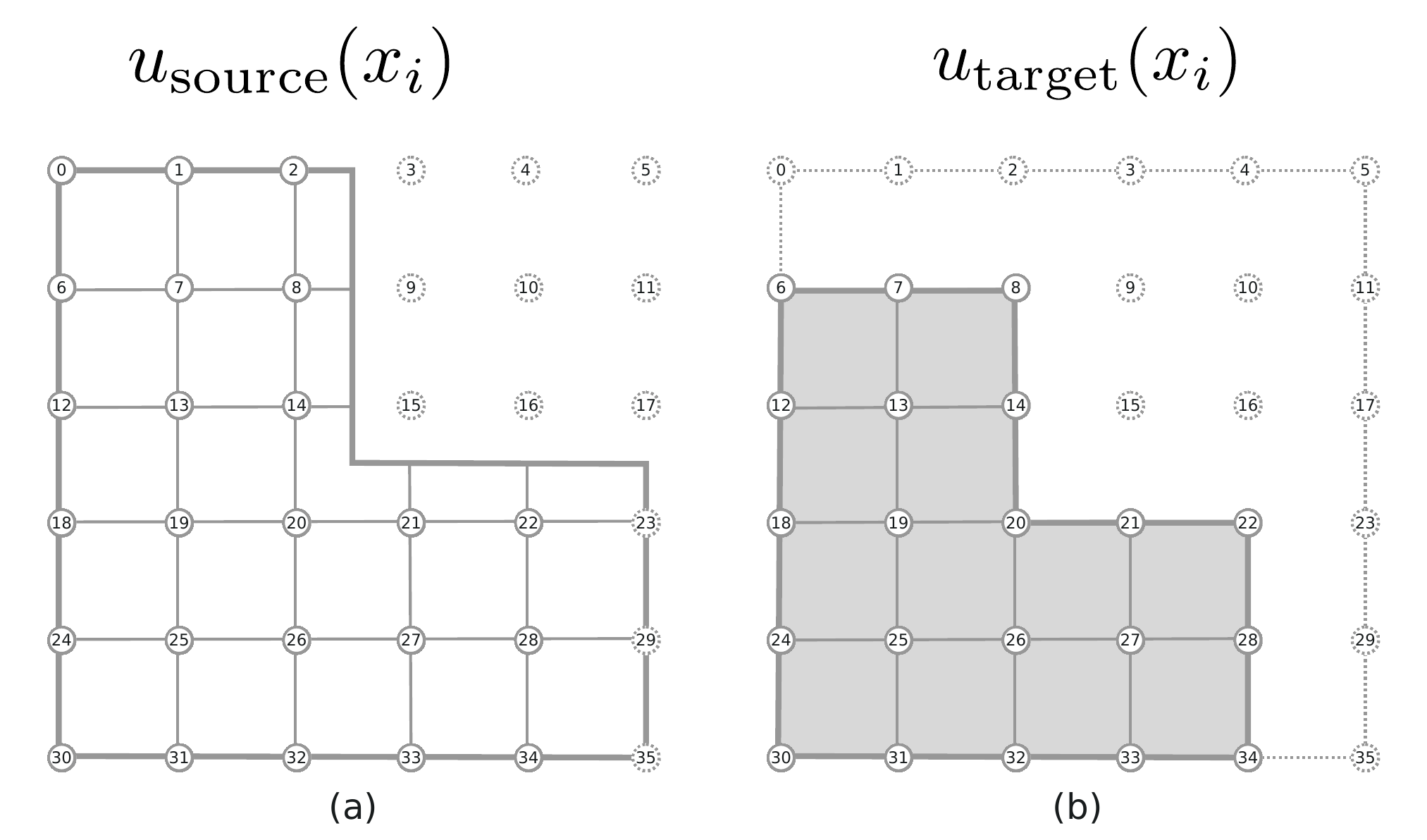}
\caption{The input function $u$ is uniformly sampled at $N$ fixed locations $x_i$ for $i=1,2,\ldots, N$ on a bounding box enclosing the geometry. The dots represent the discretization $x_i$. Sampling the input function this way facilitates transfer learning, where similar pressure values between domains are kept fixed. (a) Initial condition grid, flattened for input to the branch net as $\mathbf{u} = [x_0,x_1,\ldots,x_{35}]$, where the ghost nodes are set to zero pressures $[x_i = 0 \vert i \in \{3,4,5,9,10,11,15,16,17\}]$, $b)$ Modified initial condition grid preserving the ordering by keeping the source grid and setting the new ghost nodes to zero $[x_i = 0 \vert i \in \{0,\ldots,5,9,\ldots,11,15,\ldots,17,23,29,35\}]$. }\label{fig:input_function_transfer2D}
\end{figure}

\end{document}